\documentclass[showpacs,preprintnumbers,superscriptaddress]{revtex4}
\pdfoutput=1
\usepackage{CJK}
\usepackage{amsmath,amssymb,graphicx,bm,flushend}

\begin{document}

\title{Gravitational baryogenesis in asymptotically safe gravity}

\author{Rongjia Yang \footnote{Corresponding author}}
\email{yangrongjia@tsinghua.org.cn}
\affiliation{College of Physics Science and Technology, Hebei University, Baoding 071002, China}
\affiliation{Hebei Key Lab of Optic-Electronic Information and Materials, Hebei University, Baoding 071002, China}
\affiliation{National-Local Joint Engineering Laboratory of New Energy Photoelectric Devices, Hebei University, Baoding 071002, China}
\affiliation{Key Laboratory of High-pricision Computation and Application of Quantum Field Theory of Hebei Province, Hebei University, Baoding 071002, China}

\begin{abstract}
Gravitational baryogenesis in asymptotically safe gravity is considered. Two baryogenesis interactions are explored and their effects on the baryon-to-entropy ratio are discussed during the radiation era. The model's parameters are constrained with the current observational data, indicating these constructions may provide viable explanations for the observed baryon asymmetry.
\end{abstract}


\maketitle

\section{Introduction}
Observational evidences, such as the big-bang nucleosynthesis and the cosmic microwave background, have indicated that matter predominates over
antimatter in the universe, known as baryogenesis, which is still an important and difficult problem in both particle physics and cosmology. The baryon number of the universe from Planck 2018 data \cite{Planck:2018vyg} is determined to be of the following order
\begin{equation}
\label{1}
\eta_{\rm{B}}\equiv\frac{n_{\rm{B}}}{s}\sim(8.70 \pm 0.04)\times 10^{-11},
\end{equation}
where $s$ and $n_{\rm{B}}$ are the entropy density and the baryon number density of the present universe, respectively. Studies have suggested that the baryon asymmetry may occur during the radiation eras \cite{Riotto:1999yt, Dine:2003ax, Cline:2006ts}. Three conditions suggested by Sakharov are
necessary to generate a baryon asymmetry \cite{Sakharov:1967dj}: baryon
number violation, $\mathcal{C}$ and $\mathcal{CP}$ violation, and deviation
from thermal equilibrium. Various baryogenesis scenarios have been suggested to explain how there are more matter than antimatter in the universe. Decades years ago, a model had been proposed to generate the baryon asymmetry by violation of $\mathcal{CPT}$, while keeping
thermal equilibrium \cite{Cohen:1987vi}. A dynamical violation of $\mathcal{CP}$ was suggested to generate a baryon asymmetry in thermal equilibrium during the expansion of the universe \cite{Davoudiasl:2004gf}. The $\mathcal{CP}$-violation interaction is given by a coupling between the derivative of the Ricci scalar and the baryon current. This gravitational baryogenesis had also been investigated in $f(R)$ theories \cite{Lambiase:2006dq,Ramos:2017cot,Malakar:2025pab}, in Gauss-Bonnet gravity \cite{Odintsov:2016hgc}, in Gauss-Bonnet braneworld cosmology \cite{Bento:2005xk}, in $f(R,T)$ theories \cite{Baffou:2018hpe,Sahoo:2019pat}, in $f(T)$ theory \cite{Oikonomou:2016jjh}, in quantum fluctuation modified gravity \cite{Yang:2024gnf}, in extended teleparallel theories of gravity \cite{Bhattacharjee:2020jfk}, in $f(R, L_{\rm{m}})$ theory \cite{Jaybhaye:2023lgr}, in $f(Q, L_{\rm{m}})$ gravity \cite{Samaddar:2024qno}, in $f(T,B,T_{G},B_G)$ gravity \cite{Majeed:2025tjp}, and so on.

Recently, asymptotically safe gravity has received widespread concern. Weinberg had postulated that the effective quantum field theory of gravity might exhibit
asymptotic safety \cite{Weinberg:1980gg}: a non-Gaussian fixed point of the renormalization group flow of the couplings of the gravitational theory controls the behavior of the theory at very high energies and ensures the absence of unphysical ultraviolet divergences. This interesting idea has been applied to examine the existence of the UV fixed point in various theories, such as Einstein gravity, $f(R)$ gravity, and scalar-tensor theory (for reviews see \cite{Niedermaier:2006wt,Litim:2011cp,Percacci:2007sz}, and references therein). The effects of quantum gravity can induce modifications of Einstein equations, so it is possible that gravitational baryogenesis mechanism may occur in asymptotically safe gravity on which we focus here.

The paper is structured as follows. In Sec. II, we will briefly review the asymptotically safe cosmology. In Sec. III, we will discuss the baryon asymmetry in this scenario. Finally, we will briefly discuss and summarize our results.

\section{Asymptotically safe Cosmology}
Quantum corrections, originated from the fluctuations of quantum fields, modify the Einstein-Hilbert
action in two principal ways: additional higher-order terms in curvature must
be added to remove the divergences during the renormalization process, or the
parameters (such as $G$ and $\Lambda$) in this enlarged action acquire scale
dependence. The scale-dependent parameters can be introduced at the level of the action, at the level of the equations of motion \cite{Babic:2004ev,Shapiro:2004ch,Domazet:2010bk,Ahn:2011qt}, or at the level of solutions, if they are available in analytical forms. In this paper we concentrate on the first of these
approaches. The Einstein-Hilbert action with scale dependent couplings $G_k$ and $\Lambda_k$ is given by
\begin{equation}
\label{2}
S[g]=\int d^4 x \sqrt{-g}\left(\frac{R-2 \Lambda_k}{16 \pi G_k}+\mathcal{L}_m\right).
\end{equation}
The most general choice for the scale $k$ in Eq. \eqref{2} respecting the requirement of general
covariance is for $k$ to become a scalar filed \cite{Domazet:2012tw,Koch:2010nn,Bonanno:2012jy,Hindmarsh:2012rc}. Taking this point into mind, from Eq. \eqref{2} yields the equations of motion for the metric field
\begin{equation}
\label{eom}
G_{\mu \nu}=-g_{\mu \nu} \Lambda_k+8 \pi G_k T_{\mu \nu}-\Delta t_{\mu \nu},
\end{equation}
where $G_{\mu \nu}$ is the Einstein tensor. The possible coordinate dependence of $G_k$ leads to an additional contribution to
the stress energy tensor \cite{Carroll:2004st, Reuter:2004nv}
\begin{equation}
\Delta t_{\mu \nu}=G_k\left(g_{\mu \nu} \square-\nabla_\mu \nabla_\nu\right) \frac{1}{G_k}.
\end{equation}
Taking the covariant derivative of Eq. \eqref{eom} and assuming the conservation of
the energy-momentum tensor of matter, one derives the relation
\begin{equation}
R \nabla_\mu\left(\frac{1}{G_k}\right)-2 \nabla_\mu\left(\frac{\Lambda_k}{G_k}\right)=0.
\label{cem}
\end{equation}
Considering a Friedmann-Lema\^{i}tre-Robertson-Walker (FLRW) spacetime
\begin{align}
ds^2=-dt^2+a^2(t)\left[\frac{dr^2}{1-Kr^2}+r^2(d\theta^2+\sin^2\theta d\phi^2)\right],
\end{align}
where $a$ is the scalar factor. The spatial curvature constant $K=+1$, $0$ and $-1$ corresponds to a closed, flat, and open universe, respectively. Suggesting a perfect fluid specified by the energy-momentum tensor
\begin{eqnarray}
T_{\mu\nu}=(\rho+p)u_{\mu}u_{\nu}+g_{\mu\nu}p,
\end{eqnarray}
with $\rho$ the energy density and $p$ the pressure, then the Einstein equations reduce to the generalizations of Friedmann equations
\begin{equation}
\label{f12}
\left(\frac{\dot{a}}{a}\right)^2=\frac{8 \pi G_k}{3}\rho+\frac{\Lambda_k}{3}-\frac{K}{a^2}+\frac{\dot{G}_k \dot{a}}{G_k a},
\end{equation}
and
\begin{equation}
\label{f2}
\frac{\ddot{a}}{a}=-\frac{4 \pi G_k}{3}(1+3w)\rho+\frac{\Lambda_k}{3}+\frac{1}{2}\frac{\dot{G}_k }{G_k }\frac{\dot{a}}{a}+\frac{1}{2}\frac{\ddot{G}_k}{G_k}-\left(\frac{\dot{G}_k}{G_k}\right)^2,
\end{equation}
with the equation of state $w=p/\rho$ for matter. According to the Planck 2018 results, $\Omega_{\rm K0}=0.001\pm 0.002$ \cite{Planck:2018vyg}, here we focus on a spatial flat FRWL spacetime. Assuming that the scale factor and the running parameters $G_k$ evolve as power-law: $a(t)=a_0(t/t_0)^\alpha$, $G_k=G_0(t/t_0)^\beta$, respectively, where $\alpha$, $\beta$, $a_0$, $G_0$ and $t_0$ are constants, we find from Eqs. \eqref{f12} and \eqref{f2}
\begin{equation}
\label{f3}
\left(\frac{\dot{a}}{a}\right)^2-\frac{\ddot{a}}{a}=4 \pi(1+w) G_k\rho+\frac{1}{2}\frac{\dot{G}_k}{G_k}\frac{\dot{a}}{a}
-\frac{1}{2}\frac{\ddot{G}_k}{G_k}+\left(\frac{\dot{G}_k}{G_k}\right)^2.
\end{equation}
which gives after some calculations
\begin{equation}
\label{rhos}
\rho=\frac{1}{8\pi}\frac{(2\alpha-\beta-\alpha\beta-\beta^2)t^\beta_0}{(1+w)G_0 t^{\beta+2}},
\end{equation}
from which we have $\beta=4\alpha-2$ for radiation with $\rho=\rho_0 a^{-4}$. Supposing that in the radiation era the universe evolves slowly from an equilibrium state to an equilibrium state, the connection between energy density and temperature $T$ is given by
\begin{equation}
\label{rot}
\rho=\frac{\pi^{2}}{30} g_{*} T^{4},
\end{equation}
where $g_{*}$ denotes the number of the degrees of freedom of the effectively massless particles. Using this equation and the energy density \eqref{rhos}, we find the decoupling time as a function of decoupling temperature $T_{\rm{D}}$ as
\begin{equation}
\label{tt}
t_{\rm{D}}=
\left[\frac{15}{4}\frac{(2\alpha-\beta-\alpha\beta-\beta^2)t^\beta_0}{\pi^3 g_{*}(1+w)G_0T^4_{\rm{D}}}\right]^{\frac{1}{\beta+2}}.
\end{equation}
With this equation in gravitational baryogenesis, we can discuss whether the baryon asymmetry can be generated during the expansion of the universe in asymptotically safe gravity when a $\mathcal{CP}$-violating interaction is given.

\section{Gravitational baryogenesis}
In this section, we will consider two types of gravitational baryogenesis, One of them is newly proposed here. We will constrain the model's parameters with observational data and discuss whether the baryon asymmetry can be achieved.

\subsection{Old gravitational baryogenesis}
In the old gravitational baryogenesis, the interaction between the derivative of the Ricci scalar
$R$ and the baryon-number current $J_\mu$ breaks $\mathcal{CPT}$ dynamically in the expansion of the universe. The $\mathcal{CP}$-violating interaction term takes the form

\begin{equation}
\label{Rc}
\frac{1}{M_{*}^{2}} \int d^4 x \sqrt{-g} J^{\mu} \partial_{\mu} R,
\end{equation}
with $M_*$ the cutoff scale of underlying effective theory. The baryon-to-entropy ratio in the expansion of universe, when the temperature of the universe drops below the critical temperature $T_{\rm{D}}$, becomes \cite{Davoudiasl:2004gf}
\begin{equation}
 \label{nb}
\eta_{\rm{B}} \simeq -\frac{15 g_{\rm{b}}}{4 \pi^{2} g_{*}} \frac{\dot{R}}{M_{*}^{2}T_{\rm{D}}}.
\end{equation}
The Ricci scalar in a flat FLRW spacetime is given by
\begin{equation}
\label{rt}
R=6\left(\dot{H}+2 H^{2}\right)=-\frac{6\alpha(2\alpha-1)}{c^2t^{2}},
\end{equation}
and its time derivative is
\begin{equation}
\label{rdot}
\dot{R}=-\frac{12\alpha(2\alpha-1)}{t^{3}}=\frac{12\alpha(2\alpha-1)}{c^2t^{3}}.
\end{equation}
Combining this equation with Eq. \eqref{tt}, we derive
\begin{eqnarray}
 \label{nb1}
 \eta_{\rm{B}} \simeq\frac{45 g_{b}\alpha(2\alpha-1)}{\pi^{2} g_{*}M_{*}^{2}T_{\rm{D}}}\left[\frac{15}{4}\frac{(2\alpha-\beta-\alpha\beta-\beta^2)t^\beta_0}{\pi^3 g_{*}(1+w)G_0T^4_{\rm{D}}}\right]^{-\frac{3}{\beta+2}}.
\end{eqnarray}

If $T_{\rm{D}}=M_{\rm{I}}$ where $M_{\rm{I}}\sim 2\times10^{16}$ GeV is the upper limit on the tensor mode fluctuation constraints in inflationary scale, a possible choice of the cutoff scale $M_{*}$ is $M_{*}=m_{\rm p}/\sqrt{8\pi}$, where $m_{\rm p}$ is the Planck mass \cite{Davoudiasl:2004gf}. Taking $w=1/3$, $m_{\rm p}\sim 1.22\times 10^{19}$ GeV, $g_{\rm{b}}\simeq \mathcal{O}(1)$, $g_*\simeq 106$, $t_0\sim 1/m_{\rm p}$, and $G_0\sim 1/m^2_{\rm p}$, then \eqref{nb} reduces to

\begin{figure}
\centering
\includegraphics[height=8cm,width=10cm]{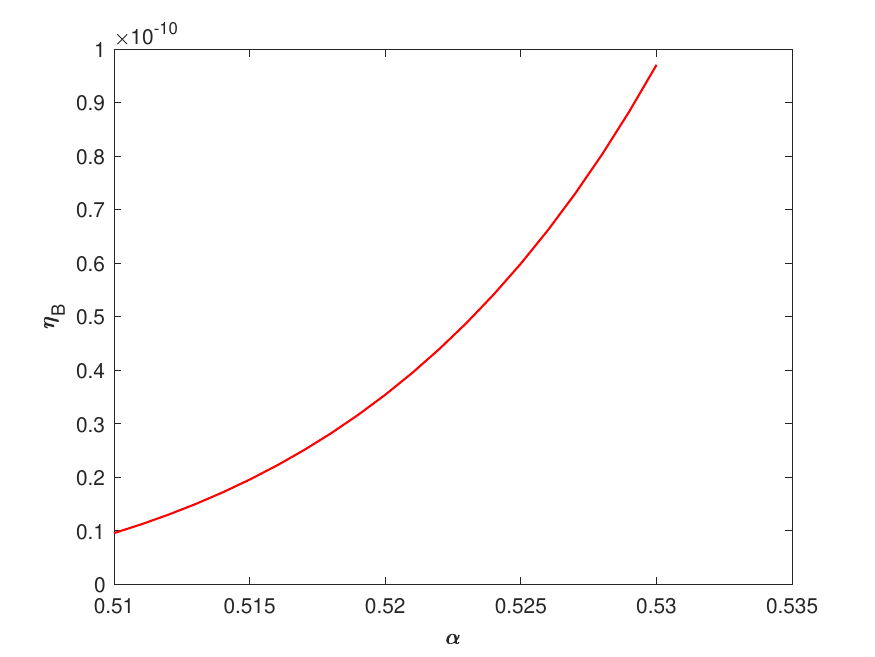}
\caption{The baryon-to-entropy ratio as a function of $\alpha$ in the coupling model $J_\mu \partial^\mu R$.}
\label{R1}
\end{figure}

\begin{eqnarray}
 \label{nb2}
\eta_{\rm{B}}\simeq\frac{360\times 610^{\frac{\beta-10}{\beta+2}} g_{b}\alpha(2\alpha-1)}{\pi g_{*}}\left[\frac{15}{4}\frac{(2\alpha-\beta-\alpha\beta-\beta^2)}{\pi^3 g_{*}(1+w)}\right]^{-\frac{3}{\beta+2}}.
\end{eqnarray}
Eq. \eqref{nb2} is very complex, we can't solve it analytically. However, we find in numerical calculations that the parameter are constrained as:
$0.51<\alpha< 0.5304$, if assuming that $n_{\rm{B}}/s\sim 10^{-11}$. In table \ref{ort1}, we present some valves of the baryon-to-entropy ratio for some given values of the parameters $\alpha$ and $\beta$.
\begin{table}
\centering
\renewcommand\arraystretch{3}
\setlength{\tabcolsep}{2mm}{
\begin{tabular}{c| c |c |c |c }
\hline
 $\alpha$ & 0.51  & 0.525 & 0.529 & 0.5304  \\
\hline
 $\beta$ & 0.04  & 0.10 & 0.115 & 0.1215\\
\hline
 $\eta_{\rm{B}}$ & $9.58\times 10^{-12}$  & $6.00\times 10^{-11}$ & $8.63\times 10^{-11}$ & $1.00\times 10^{-10}$\\
\hline
\end{tabular}}
\caption{Some values of the baryon-to-entropy ratio for some given values of the parameters $\alpha$ and $\beta$ in the coupling model $J_\mu \partial^\mu R$.
}
\label{ort1}
\end{table}

In figure \ref{R1}, we visualize the variation of baryon-to-entropy ratio with the parameter $\alpha$. From the graph, we observe that the baryon-to-entropy ratio increases with the increase of the parameter $\alpha$. This implies that a lower value of $\alpha$ suppresses the baryon asymmetry generated from old
gravitational baryogenesis. Figure \ref{R1} also shows that the baryogenesis can be realized with suitable parameter spaces, producing a net baryon asymmetry factor consistent with observations.

\subsection{General gravitational baryogenesis}

\begin{figure}
\centering
\includegraphics[height=8cm,width=10cm]{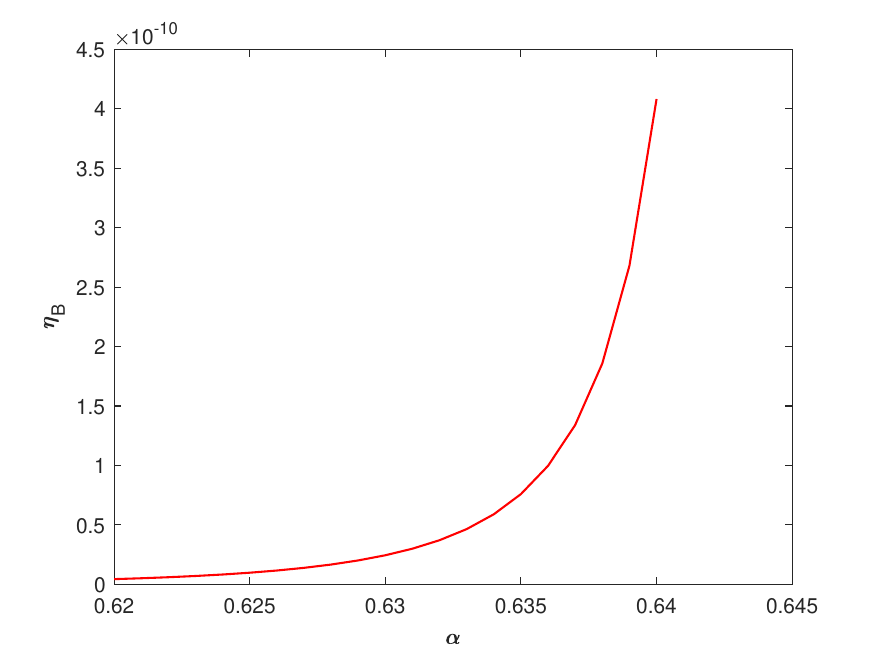}
\caption{The baryon-to-entropy ratio as a function of $\alpha$ in the coupling model $J_\mu \partial^\mu R^2$.}
\label{nr}
\end{figure}

In literatures, the interaction term $J_\mu\partial^\mu R$ was usually discussed. Considering other possibilities is valuable in the absence of the fundamental principle of choosing the $\mathcal{CP}$-violating interaction term. We consider a general $\mathcal{CP}$-violating interaction term, and then discuss a specific case in order to limit the parameters with observational data. Taking a general $\mathcal{CP}$-violating interaction term
\begin{equation}
\frac{1}{M^2_*}\int d^4 x \sqrt{-g} J^{\mu} \partial_{\mu} f(R),
\end{equation}
where $f(R)$ is any function of $R$, the baryon-to-entropy ratio can be calculated as
\begin{equation}
\label{nr2}
\eta_{\rm{B}}\simeq -\frac{15 g_{\rm{b}}}{4 \pi^{2} g_{*}} \frac{\dot{R}f_{R}}{M^2_*T_{\rm{D}}},
\end{equation}
where $f_{R}\equiv df/dR$. Obviously, the value of $\eta_{\rm{B}}$ depends on the specific form of $f_{R}$. As a specific case, thinking about: $f(R)=R^2/M^2_{*}$, then from Eq. (\ref{nr2}), yields
\begin{eqnarray}
\label{nr21}
\eta_{\rm{B}}\simeq -\frac{30 g_{\rm{b}}}{4 \pi^{2} g_{*}}\frac{R\dot{R}}{M_{*}^{4}T_{\rm{D}}}.\\
\end{eqnarray}
Inserting Eqs. \eqref{tt}, \eqref{rt}, and \eqref{rdot} into this equation and taking the same values for the parameters as above, after some calculations we find
\begin{eqnarray}
\label{nr211}
&&\frac{n_{\rm{B}}}{s}\simeq \frac{2.10816\times10^7 \times610^{-\frac{20}{\beta+2}}g_{\rm{b}}\alpha^2(2\alpha-1)^2}{g_{*}}
\left[\frac{15}{4}\frac{(2\alpha-\beta-\alpha\beta-\beta^2)t^\beta_0}{\pi^3g_{*}(1+w)G_0T^4_{\rm{D}}}\right]^{-\frac{5}{\beta+2}}.
\end{eqnarray}

 Eq. \eqref{nr211} is also very complicated, we cannot analytically solve for the values of the parameters. In numerical calculations, however, we find that the parameters are constrained as: $0.6251<\alpha<0.6361$. Some values of the baryon-to-entropy ratio for some given values of the parameters $\alpha$ and $\beta$ are presented in table \ref{ort2}.

 \begin{table}
\centering
\renewcommand\arraystretch{3}
\setlength{\tabcolsep}{2mm}{
\begin{tabular}{c| c |c |c |c }
\hline
 $\alpha$ & 0.6251  & 0.63 & 0.635 & 0.6361 \\
\hline
 $\beta$ & 0.5004  & 0.52 & 0.54 & 0.5444 \\
\hline
 $\eta_{\rm{B}}$ & $9.98\times 10^{-12}$  & $2.45\times 10^{-11}$ & $7.59\times 10^{-11}$ & $1.03\times 10^{-10}$\\
\hline
\end{tabular}}
\caption{Some values of the baryon-to-entropy ratio for some given values of the parameters $\alpha$ and $\beta$ in the coupling model $J_\mu \partial^\mu R^2$.
}
\label{ort2}
\end{table}

We visualize the baryon-to-entropy ratio as a function of parameter $\alpha$ in figure \ref{nr}. We observe that the baryon-to-entropy ratio also increases when the parameter $\alpha$ increases in this model, meaning that larger value of $\alpha$ will enhance the baryon asymmetry. From the graph, a net baryon asymmetry factor consistent with observations can be produced with suitable parameter spaces, meaning that he baryogenesis can be realized.

\section{Conclusions and discussions}
We have considered the gravitational baryogenesis in asymptotically safe gravity. First, we have discussed the old gravitational baryogenesis and explored the effect of this interaction term on the baryon-to-entropy ratio during the radiation era of the expanding universe. Second, we have introduced a general gravitation induced $\mathcal{CP}$-violating term and discussed a specific case to investigated whether the baryon asymmetry can be realized. The calculations yielded values for the baryon-to-entropy ratio that are in agreement with observational data, reaching an order of magnitude around $10^{-11}$. These results confirm that the gravitational baryogenesis mechanism within asymptotically safe gravity can successfully explain the matter-antimatter asymmetry observed in the universe

\begin{acknowledgments}
This study is supported in part by National Natural Science Foundation of China (Grant No. 12333008).
\end{acknowledgments}
\bibliographystyle{elsarticle-num}
\bibliography{refq}

\end{document}